\documentclass{article}

\usepackage{amssymb,amsfonts,amsmath}
  \usepackage{ccaption}
  
 \usepackage[dvips]{graphicx}

\usepackage{amssymb,amsfonts,amsmath}

\newcommand{\react}[1]{\stackrel{#1}{\rightarrow}}

\def\reactionrates#1#2{\mathrel{\mathop{\rightleftharpoons}\limits^{#1}_{#2}}}

\usepackage[blocks]{authblk}

 \captionnamefont{\bfseries}
  \captiontitlefont{\small\sffamily}

\begin{document}

\title{Resonant activation: a strategy \\against bacterial persistence}\date{}
\author[1,2]{Yan Fu}
\author[3]{Meng Zhu}
\author[2,*]{Jianhua Xing}

\affil[1]{Interdisciplinary Program of Genetics, Bioinformatics, and Computational Biology, Virginia Polytechnic Institute and State University, Blacksburg, VA 24060}
\affil[2]{Department of Biological Sciences, Virginia Polytechnic Institute and State University, Blacksburg, VA 24060}
\affil[3]{School of Computing, Clemson University, Clemson, SC 29631}
\affil[*]{Corresponding author: jxing@vt.edu}

\pagestyle{empty}
\maketitle

Short title: Resonant activation against bacterial persistence


\begin{abstract} 
A  bacterial colony may develop a small number of cells genetically identical to, but phenotypically different from other normally growing bacteria. These so-called persister cells keep themselves in a dormant state and thus are insensitive to antibiotic treatment, resulting in serious problems of drug resistance. In this paper, we proposed a novel strategy to ``kill'' persister cells by triggering them to switch, in a fast and synchronized way, into normally growing cells that are susceptible to antibiotics. The strategy is based on resonant activation (RA), a well-studied phenomenon in physics where the internal noise of a system can constructively facilitate fast and synchronized barrier crossings. Through stochastic Gilliespie simulation with a generic toggle switch model, we demonstrated that RA exists in the phenotypic switching of a single bacterium. Further, by coupling single cell level and population level simulations, we showed that with RA, one can greatly reduce the time and total amount of antibiotics needed to sterilize a bacterial population. We suggest that resonant activation is a general phenomenon in phenotypic transition, and can find other applications such as cancer therapy.
\end{abstract}

\clearpage

\clearpage
\section{Introduction}
Noise has often been viewed as a nuisance for many years in biology. Robustness of large biological systems requires noise from both intracelluar and intercelluar sources being canceled or filtered in one way or another. Yet growing evidence indicates that noise actually plays fundamental roles in many biological processes, as in cell fate decision and in mutation and evolution \cite{Rao2002}. In this paper, we suggest that the functional role of noise can go even beyond: the internal noise in bacterial gene expression can be utilized to counteract antibiotic resistance, by inducing resonant activation that can facilitate a fast and synchronized phenotypic switching in bacteria population.

Antibiotic resistance is a severe and growing problem in clinical practice. It refers to a phenomenon that certain phenotypes of microorganisms, {\it e.g.} bacteria, are able to withstand (and requires prolonged) antibiotic treatment.  It may be acquired from horizontal gene transfer and mutations in the pathogenic chromosome \cite{Ochman2000, Cirz2005, Miller2004}, or from the existence of phenotypic heterogeneity within bacteria population \cite{Balaban2004, Levin2006,Gefen2009}. The latter links antibiotic resistance to a special bacterial phenotype called persister cell, a non-growing (or slowly growing) and non-inherited cell phenotype whose number only accounts for a small fraction of total population. 

Persister cells are genetically homogeneous to normally growing cells. They are first discovered by Bigger \cite{Bigger1944} and then found in higher-percentage in biofilms that are known to be responsible for a majority of recalcitrant infections such as tuberculosis\cite{Ojha2008,Stewart2003}. Several experimental and theoretical works suggested their formation during mid-exponential phase, as well as their function as an ``insurance" to maximize the overall survival probability of bacterial population in changing environments \cite{Balaban2004, Keren2004a,  Kussell2005, Lou2008}. The magic is that persister cells have negligible growth rate and non-negligible phenotypic switching rate. The negligible growth rate helps persister cells dodging antibiotic attack which depends mostly on active cell wall growth. The non-negligible phenotypic switching rate, on the other hand, ensures finite probability of stochastic switching from persister cells to normally growing cells taking place at a period that the stress (e.g., antibiotics) is removed. Then those newly formed normal cells serve as the ``seeds" for reestablishing the population.  Therefore, Kussell {\it et al.} proposed the phenotypic switching rate can be seen as a result of evolutionary adaptation of bacteria to their real fluctuating environment \cite{Kussell2005} . 

The ubiquity of persister cells makes bacteria population hard to sterilize. The time-series of the survival fraction under antibiotic treatment obeys a two-phase exponential decay, with the majority of cells being killed at a fast rate at the beginning while the rest being killed much slowly afterwards. {\it hipA7}, a mutant strain of {\it E.coli} that contains higher percentage ($10^{-5}\sim10^{-2}$) of persister cells, has been reported survival in a fraction about $10^{-5}$ even after continuous ampicillin treatment for 50 hrs \cite{Balaban2004}. Therefore, It is of concern how to efficiently sterilize bacteria populations, especially for strains with more fraction of persister cells. In 2008, Gefen {\it et al.}   observed that persister cells of {\it hipA7} assume normal growth during the first 1.5 hour on exiting the stationary phase \cite{Gefen2008}. While applying ampicillin within that time window, the number of persister cells can be lowered by 1 order of magnitude. However, those still alive may adopt dormancy after that time window, and can convert to normally growing phenotype and re-grow to a new population under appropriate conditions. The essential problem here is that well-established persister cells are insensitive to antibiotics. They have to be converted into normally growing cells to get sterilized by drugs. However this transition is stochastic and may take a long time. For {\it hipA7}, the transition rate is $0.07hr^{-1}$, which gives an exponential waiting time distribution with the  average waiting time of conversion $\sim 14 \ hr$.  For some species, {\it e.g.} the {\it E coli} mutant {\it hipQ}, the rate can be much smaller. Therefore, bacteria sterilization requires continuous antibiotics application at least to cover this broad range of time, which may be impractical and/or detrimental to the host. 

Several toxin-antitoxin (TA) modules in bacterial chromosome have been experimentally identified to regulate bacterial phenotypic transitions, though detailed mechanism has not been clearly understood\cite{Gefen2009, Pedersen2002, Christensen2003}. Irrespective of the details, the basic mechanism is simple: a mutual inhibition exists between antitoxin's and cognate toxin's expression, which determines whether a single bacterium assumes normally growing phenotype (when antitoxin dominates) or persister phenotype (when toxin dominates). Therefore bistability is the major  dynamical property of single cell's phenotypic transition\cite{Lou2008}. In this theoretical investigation we used the well studied generic toggle switch to model this bistable system, Figure 1 (a).  

Making analogy between biological networks and other familiar physical systems has lead to several insightful studies \cite{Zhu2004, Walczak2005, Wang2006}. In this work we focused on resonant activation (RA), a well studied phenomenon for thermally activated barrier-crossing systems \cite{Schmitt2006, Doering1992, Marchi1996}. If the barrier is under time varying periodic perturbation, ``cooperative interplay between the barrier modulation process and thermal noise assisting barrier crossing events can cause an enhancement of the reaction kinetics" \cite{Schmitt2006}. The mean first passage time (MFPT), which is the average time the system waits for the first successful barrier crossing, reaches its minimum (by several folds or even orders of magnitude compared with that of the unperturbed system) at the resonance frequency of the perturbation. Under resonance frequency $\omega_c$ with period $T_{\omega_c} = 2\pi/\omega_c$, the system prefers to make a transition when the barrier height reaches its minimum. Consequently,  the FPT distribution displaces a series of peaks at odd multiples of $T_{\omega_c}/2$, instead of a continuous exponential distribution.  RA is related but different from another well studied phenomenon, stochastic resonance \cite{Schmitt2006, Gammaitoni1998}. 

Complimentary to current efforts of searching for more efficient antibiotics, in this work we propose to utilize the phenomenon of resonant activation to help fighting drug resistance. Noticing the similarity between thermally activated barrier crossing and cell phenotypic transition, we will first use a genetic toggle switch model to demonstrate that resonant activation exists for biological network dynamics. Then we will examine how one can shorten the time and the amount of antibiotics needed to extinct a bacteria population using resonant activation. Our strategy utilizes the two characters of RA: accelerated kinetics, and multi-peaked FPT distributions.

\section{Model and Method}
\subsection{Single cell level}
Currently the exact regulation mechanism for the persister-normal cell transition is not clear, and different hypotheses have been raised to give possible answers\cite{Gefen2009}. For our purpose of illustrating the idea, we will use a generic toggle switch to represent the mutual inhibition within T-A module. As shown in Figure 1 (a), the network contains two genes mutually inhibiting each other through their dimerized protein products. The two types of protein dimers compete for the promotor binding site. Ten chemical reactions shown below control the dynamics of the generic toggle switch shown in Figure 1 (a).  \\
$O\react{k_{1}^0} O+A$, $O\react{k_{2}^0} O+B$\\
$A\react{k_{3}^0}\phi$, $B\react{k_{4}^0}\phi$\\
$A+A\reactionrates{k_5}{k_{-5}} A_2$, $B+B\reactionrates{k_6}{k_{-6}} B_2$ \\
$O+A_2\reactionrates{k_7}{k_{-7}}OA_2$, $O+B_2\reactionrates{k_8}{k_{-8}}OB_2$\\
$OA_2\react{k_{9}} OA_2+A$, $OB_2\react{k_{10}} OB_2+B$\\
Table 1 gives the corresponding rate constants in reduced unit. With these parameters and mass-action type dynamics, the system gives two stable states corresponding to the persister (the difference between the total number of free and bound protein B and the total number of free and bound protein A, denoted as $\delta$, is less than 0), and normally growing ($\delta>0$) phenotypes, respectively (see Figure 1 (b) ). Stochastic fluctuations of the reactions drive the system transit between the two states. These parameters are modified from the model of Allen {\it et al.} \cite{ Allen2005}, so the model gives  {\it hipA7} $p2n$ (persister cell to normally growing cell)   and $n2p$ (normally growing cell to persister cell) switching rates $0.07/hr$ and $0.008/hr$ (with time unit $1 \ unit=130 \ hr^{-1}$), respectively, as used by others  \cite {Balaban2004, Kussell2005}.   The abstract toggle switch model also places the present work in a broader context. Many microorganisms, including viruses and bacteria, coexist in a dormant and an active phenotype \cite{Dubnau2006}. The toggle switch is a frequently occured generic regulation mechanism and a good model for phenotypic transitions in bacteria\cite{Smits2006, Gardner2000}.

To be specific and for practical considerations of the computational feasibility, we chose model parameters in most simulations in this work to mimic the dynamics of {\it hipA7} . However, we want to emphasize that the proposed approach below works best for the following situation. First,  we assume that one can regulate some of the rate constants through an external oscillating perturbing signal. Consequently some rates (the degradation rate of A for the results reported in the main text) are oscillating with time. The strength of the perturbing signal should be restricted due to the consideration of toxicity to the host.   Second, we focus on the case that the transition from the persister to the normally growing phenotype is very slow, so that the sterilization of the total population requires a long term antibiotic treatment, which may bring severe side-effect like liver damage.

\subsection{Population level}
At the population level, a cell is subject to an environment alternating between growing and antibiotics stress conditions. A normally growing cell has faster net proliferation rate than a persister cell does under growth condition, but also a larger death rate under antibiotic stress (see also \cite{Kussell2005}). 
Three types of cellular events can take place for each cell: cell division--the cell including the molecular state of the goggle switch are cloned into two identical  copies, cell death, and phenotypic transition. Table 2 gives all the related rate constants based on experimental observations \cite {Balaban2004, Kussell2005}. For a given cell with phenotype $n$ (normally growing) or $p$ (persister), the reactions for stochastic simulations additional to the 10 toggle switch reactions are,\\
$n\react{g_n}2n$, $n\react{d_n}\phi$, $p\react{g_p}2p$, $p\react{d_p}\phi$. \\
Unlike the work of Kussell {\it et al}, in our model we did not simulate the cell phenotypic transitions directly. Instead we propagate the 10 toggle switch reactions for each cell, which determine the phenotype of the cell (see above). 
Each population level simulation initiates from a stationary-phase colony including $10^4$ normal cells and $10^2$ persister cells. The population is then put into fresh medium with/without antibiotics, as determined by each different strategy.  The phenotype of a single cell is determined by its own  toggle switch dynamics based on the value of $\delta$. We use $\tau$-leap Gillespie algorithm to propagate the 10 chemical reactions of each toggle switch and cell birth/death simultaneously \cite{Gillespie2001} .  For the rates with periodic time-dependence, we approximate them as constant within one Gillespie step, which is much smaller than the rate varying period.  For simplicity, we do not consider quorum sensing, thus each cell behaves independently and does not communicate with others except for competing resources as discussed below. To prevent the population from overgrowth in our simulations, we rescaled the growth rate $g$ as a decreasing function of the  number of normally growing cells $n$, $g_n(t+\tau)={g_n^0} /[{1+\alpha\cdot n(t)]}$ where $\tau$ is the time step in $\tau$-leap Gillespie algorithm, $\alpha$ controls the scaling strength (here $\alpha=0.001$), and $g_n^0$ represents the original growth rate of normal cells without rescaling. We didn't rescale the persister cells' growth rate since the value before rescaling is already negligible. A physical justification of the rescaling is that  the accessible nutrients and volume of a cell colony are usually limited against unrestricted massive replications. 

\section{Results}

\subsection{Resonant activation exists in cellular phenotypic switch}
For the model we examined, the switching rate from $n2p$  is much smaller than that of $p2n$ at the exponential phase. This dynamics mimics that of {\it hipA7}  \cite{Balaban2004}, and resembles to barrier crossing rates in an asymmetric double well potential, with one deeper well representing normally growing phenotype, and the other well representing persister phenotype. Let's define $t_{p2n}$ as the first passage time for $p2n$ switching. The distribution of $t_{p2n}$, apart from an initial  transient time period, follows an exponential form $$P(t_{p2n}) \propto exp(-t_{p2n}/t_K)$$ (see Figure 2 (a)), where $t_K$ is the inverse of the Kramers rate \cite{Gammaitoni1998, Hanggi1990, Pechukas1994}. The prolonged distribution contributes to bacterial persistence.

Next, we perturbed the protein $A$'s degradation rate with sine-formed signal (see Figure 1), $k_3 = k_3^0 [1+ \gamma\sin(\omega t +\theta)]$, with $0<\gamma<1$. For each simulation the phase $\theta$ is randomly drawn from a uniform distribution betwen $0$ and $2\pi$. This is because in the population level simulations below, the relative phase between  the birth time of a persister cell and the added signal can be seen as a random variable. Figure 2 (b) shows that the mean first passage time (MFPT) as a function of $\omega$, averaged over 5000 independent simulations, shows a minimum around $\omega_c=1.6 \ hr^{-1}$, although the curve is rather flat over a range of $\omega$ values. Existence of the minimal MFPT signatures resonant activation \cite{Schmitt2006, Marchi1996}, despite the current system is described by discrete dynamics.

Figure 2 also shows $P(t_{p2n})$ under the perturbing signal with different frequency. At a high $\omega$, the system cannot respond fast enough, and the perturbation is equivalent to an averaged constant one. The distribution is exponential, Figure 2 (c).  Under resonance frequency $\omega_c$, however, $t_{p2n}$ distribution changes into several separated spikes, Figure 2 (d). Note that  the peaks of spikes overlap with the peaks of the periodic signal. Therefore transition takes place more frequently  when the signal reaches the peak value, thus the degradation rate of protein A is the fastest, and the transition from A dominant to B dominant is the easiest  \cite{Gammaitoni1998}. If the system misses one peak of the signal for a {\it p2n} transition, it prefers waiting for the next peak. We observed this type of localized distribution over a broad range of $\omega$ values corresponding to the flat bottom region of the MFPT-$\omega$ curve (see Figure 2 (b) and Figure 6 in the Supplementary Information). Further increase of $\omega$ leads to gradual merge of the spikes, and eventually reduce to a single exponential distribution. These observations are consistent with studies with barrier crossings in a double well potential \cite{Schmitt2006}, further supporting the existence of RA in the system.

For all the results reported here, we added the perturbing signal on protein $A$'s degradation for illustrative purpose (see Figure 1 (a) ). In real experiments and applications,  one can also perturb other reactions ({\it e.g.} protein synthesis as well as degradation), depending on the actual practical feasibility. For example, one possible implementation of this perturbation may be through varying the activity of protease through specific regulating molecules. We further presented results with perturbations on A's synthesis rate, and B's synthesis and degradation rates in the Supplementary Information, Figure 7. In all these cases we observed resonant activation with the same resonant frequency, but with varying fold of change of the MFPT. We provided a theoretical explanation there. It depends on the system to identify the reactions most sensitive to the perturbations. Figure 8 in the Supplementary Information also showed that existence of RA is independent of the detailed form of the periodic perturbation signal.  

\subsection{Resonant activation accelerates bacteria colony sterilization}
The stochastic simulations on single cell dynamics discussed above show that resonant activation can facilitate fast and synchronized $p2n$ switches. Next we coupled the single cell level dynamics with population level proliferation/death under changing environment.
 
First  we define a killing strategy K to be the combination of a perturbing signal $S$ and an antibiotic environment $E$, characterized by their adding frequency ($\omega_1$ and $\omega_2$), strength, and duration, respectively. 

For a given population initialized with $10^4$ normally growing cells and $10^2$ persister cells, our simulations allow it to evolve until no bacterium exists, or a maximum time reaches. Because the stochastic nature of the dynamics, for each population the sterilization time $T_{kill}$ is random. To quantitatively compare different strategies, for each strategy we performed  independent simulations with 1000 populations, and recorded $T_{kill}$ of each population sample.  Figure 9 in the Supplementary Information gives the killing time distribution of 1000 population samples under strategies K1 and  K3. To compare the different strategies, we used the time needed to sterilize 90\% of the 1000 population samples, $Q$, as a criterion. Practically this is a more relevant quantity than the average killing  time, although we reached similar conclusion below with the latter. 

The stochastic simulation results are summarized in Figure 3 and Table 1.  Figure 4 give several typical trajectories. Without the perturbing signal, both the strategies with periodic (K1) and continuous (K2) antibiotics treatment require long time, since it takes long time to eradicate the persister cells. On the other hand, with the perturbation at the resonance frequency (K3 an K4), the sterilization time is greatly reduced. Because under K3, most of the $p2n$ transitions take place within the period of applying antibiotics,  it is difficult for the bacteria population to restore either the normally growing or the persister subpopulations (see Figure 10 in the Supplementary Information). Consequently the sterilization time for K3 and K4 are about the same. To further prove this, we compute over 500 independent samples the ratio ($R_{p2n}^g$) between the number of $p2n$ transitions that happen during growth environment and the total number of $p2n$ transitions. Larger $R_{p2n}^g$ corresponds to the inefficiency of the periodic antibiotic strategy, because the population may be easier to get re-estabished during growth period.  Under K3,  $R_{p2n}^g=4.3\%$, while under K1, $R_{p2n}^g \sim 46.8\%$. We observed similar efficient bacterial eradication with signal frequencies away from $\omega_c$ but still laying near the flat bottom of the MFPT~$\omega$ curve in Figure 1 (b). However, further change of the frequency (K5 and K6) shows less improvement over that of K1 or K2.  Figure 3 (b) gives the total amount of antibiotics used for each case. Compared to K2, K3 requires less than half of the time, with $\sim 25\%$ of the total amount of antibiotics.  In the Supplementary Information, we also examined how the killing time depends on the signal duration and strength, Figure S11 and Figure 12.

To further examine the strength and limitation of our proposal, we examined a more difficult case of bacterial persistence. The model is similar to what discussed above, except that the $p2n$ transition rate is reduced by a factor of 5. Correspondingly, we extend each period of alternative antibiotics and freely growing environment. This leaves more time for the system to recover under the latter environment. Figure 5 gives several typical trajectories of population level simulations. For strategy K1, we had difficulty to observe population extinction even after $10^4$ hours. While almost all the normally growing cells are killed, and the persister subpopulation size is reduced under an antibiotics environment, both subpopulations are restored to their original levels under next freely growing period. For strategy K2, the $90\%$ quantile of sterilization time for 1000 populations is 362 hrs. In comparison, the $90\%$ quantile of sterilization time for K3 and K4 are 358 hrs and 168 hrs, respectively. Though under K3, the ratio of $p2n$ transitions within growth period $R_{p2n}^g=4.3\%$, in this case a single $p2n$ transition under the freely growing environment is sufficient to restore both the two subpopulations. That explains why K3 requires almost the same sterilization time with K2. However, K3 uses about half of the amount of antibiotics needed under K2. In addition, since antibiotics are applied periodically, K3 might be better than K4 considering it leaves time for the host to recover from possible side effects of the antibiotics treatment. Therefore, in this case, one has to make a compromise between short sterilization time and side-effect reduction.
\section{Discussions and concluding remarks}

In the past decade, bacterial persistence became a spotlight in mircobiological arena. The leading actors, called persister cells, are some special dormant cells which account for only a small fraction of the total population. However, they are insensitive to antimicrobial therapy, and are able to switch back into normally growing phenotype to  initiate population regrowth. The phenotypic switching property of bacteria has been experimentally identified to be regulated by several toxin-antitoxin (TA) modules within bacterium chromosome,  such as {\it hipBA}, {\it relBE} and {\it chpA}. For example, Keren {\it et al.} reported that {\it hipBA}-knocked-out {\it E. coli} biofilm produced 150-folder-fewer persister cells under mitomycin treatment \cite{Keren2004b}. Continuous efforts have been made to reveal the detailed molecular regulation mechanism \cite{Pedersen2002, Christensen2003}.

Now, it is believed that persister cells may evolve into maximizing the overall survival possibility of bacteria population in real fluctuation environment. Many lab experiments have proved that the existence of persister cells in biofilms is responsible for many recalcitrant diseases, such as human tuberculosis, an infecious disease caused by {\it Mycobacterium tuberculosis} biofilms. The typical antibiotic treatment of this disease is as long as 6-9 months \cite{Ojha2008}. Therefore, besides the inefficiency of the therapy, the side-effect from such a long-term use of antibiotics is of serious concern.  Active research is undertaken to fight against bacteria persistence through accelerating the phenotype transition rate with chemical or physical method \cite{Gefen2009, Stewart2003}. In this work we assume such a mechanism exists, and focused on the optimal strategy to combine it with antibiotics treatment. 

Stochastic resonance has been discovered in many biological systems \cite{Hanggi2002}. Similar to Schmitt {\it et al.}, we also observed stochastic resonance in the present system. Resonant activation, on the other hand, is seldom discussed in the biology context. The phenomenon of RA is related to fluctuation resonance previously reported \cite{Lipan2005, Mettetal2008}. For a stochastic system with one steady state driven by an oscillating perturbation, there may exist a resonant frequency of the perturbation so that the system shows largest fluctuation amplitude. This is called fluctuation resonance, which resembles the peak of the energy absorption spectrum of a system. For a system with multiple steady states, larger fluctuations lead to faster transition rate to a new steady state. This is resonant activation.  Here we demonstrated the existence of RA in non-thermal systems governed by discrete chemical reaction dynamics. RA has two unique properties: reduced mean first passage time, and localized transition in time. We propose to utilize the two properties of RA to help on eliminating persistent bacteria.  As previously mentioned, the essential reason for bacterial persistence is the large time-scale of $t_{p2n}$ distribution. This requires continuous antibiotic treatment to cover most of the time period to prevent bacteria population re-establishment.  With RA, the distribution is narrowed through reduction of the MFPT. This is our main argument for using RA against bacteria persistence. Furthermore for some cases the localized spike shaped transition time distribution may allow dividing the antibiotics treatment into sessions without serious problem of bacteria population restoration. This is of special advantage by minimizing side effects of antibiotics treatment to patients if the treatment has to be long. On the other hand, our simulations show that even if one can accelerate the bacteria phenotype switching rate, improper procedure (strategy K5) leads to no improvement in fighting bacteria persistence.

In this work, we presented the general idea of modulating cellular phenotype switching through resonant activation. It should be viewed as illustrative. More detailed modeling and experimental studies are necessary to examine the feasibility and the optimal strategy for each specific system.   Detailed molecular mechanism of the toxin-antitoxin module, its interaction with related signal transduction and metabolic  pathways  should be carefully considered. The detailed model will provide information on the strategies of adding the perturbing signal. In a  bacteria colony, several persistent phenotypes may coexist. In this case a more efficient strategy may be to apply antibiotics continuously for a period, then switch to the strategy we propose here for the most persistent phenotype. The broad range of resonance frequency of one phenotype (see Figure 2 (b)) may also allows one to choose an overlapping frequency for all the phenotypes. The perturbing signal is not limited to chemicals, but any external environmental change that can affect the phenotype switching dynamics.

We want to point out that the actual performance of each strategy depends on the property of the system, especially the unperturbed $p2n$ transition time and the system noise level. Here we only focused on illustrating the basic idea, and the choice of the system is partly restricted by computational considerations. Orders of magnitude reduction of the MFPT with RA have been reported for some physical systems \cite{Boguna1998}. One may expect similar result for some phenotypic transitions.

While here we focused on bacterial persistence, we want to emphasize that resonant activation is a general phenomenon for phenotypic transitions, which are analogous to thermally activated barrier crossing processes. One may find application of the idea discussed here to other problems. For example, Spencer {\it et al.} have shown that cancer cells have persistence behavior similar to bacteria \cite{Spencer2009}. Radiotherapy is  a standard cancer treatment option. It normally consists of multi-session low dose of radiation in a few weeks. The radiation induces  cell DNA damage, which eventually leads to apoptosis. In this case it is even easier to apply RA. Here radiation is the oscillating signal, apoptosis plays the role of antibiotics, and population restoration during the treatment intervals is not a serious problem. An optimal  strategy may exist on performing the treatment utilizing resonant activation. Similar argument applies to chemotherapy. In this case stochastic resonance does not exist since the system is not bistable. 

\section{Acknowledgements.} We would like to thank Dr.Nathalie Q. Balaban for discussion in the experimental works in bacterial persistence,  Dr. John J. Tyson, Dr. Katherine C. Chen, Dr. Tongli Zhang, Dr. Vlad Elgart, Zhanghan Wu and Chun Chen for discussions and comments on this paper. We would also like to thank Dr.Yang Cao for useful discussion on stochastic simulation method.


\clearpage
\begin{figure}
 \centerline{\includegraphics[width=87mm]{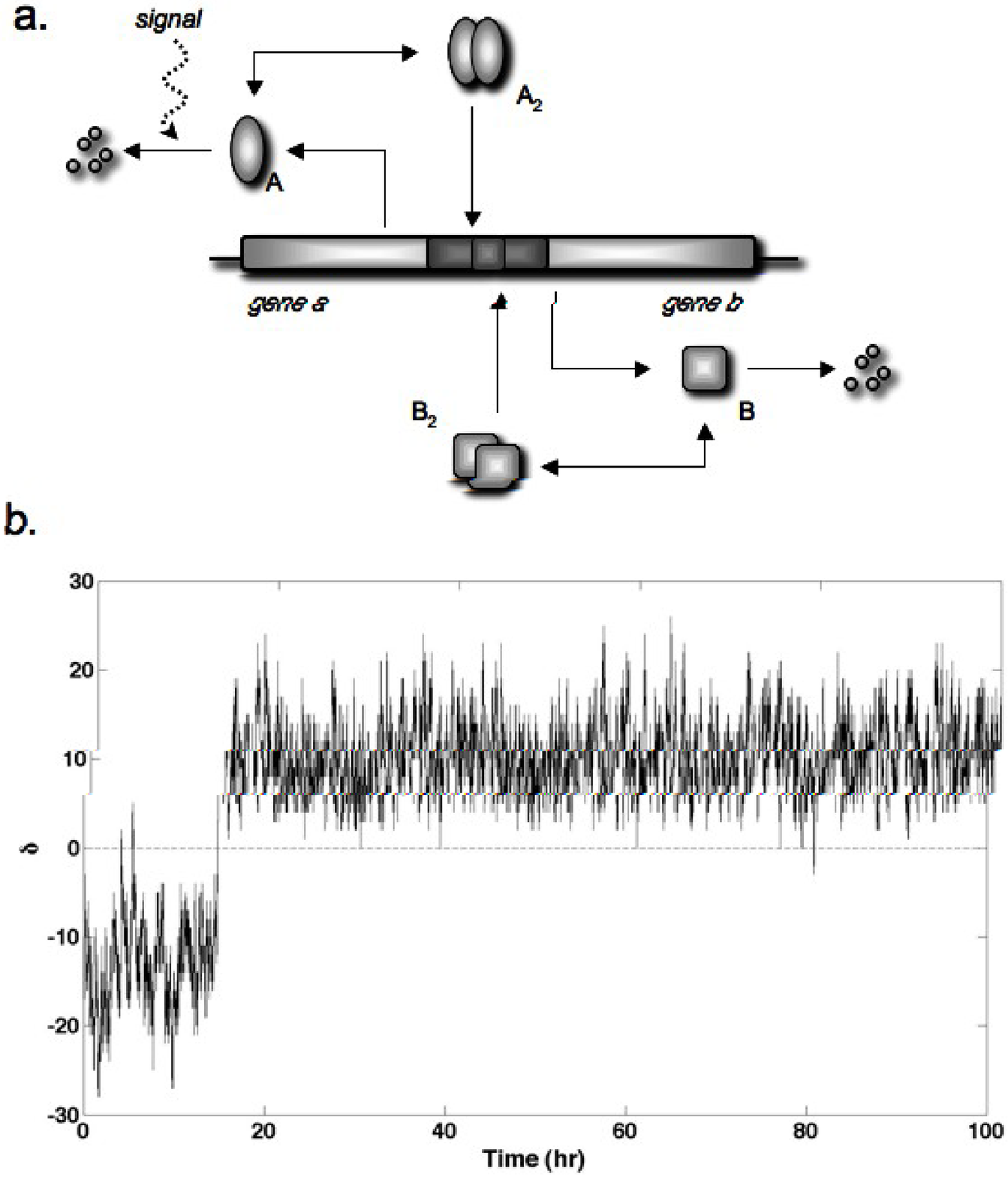}}
\caption{
 \label{Figure1}}
\end{figure}

\begin{figure}
 \centerline{\includegraphics[width=100mm]{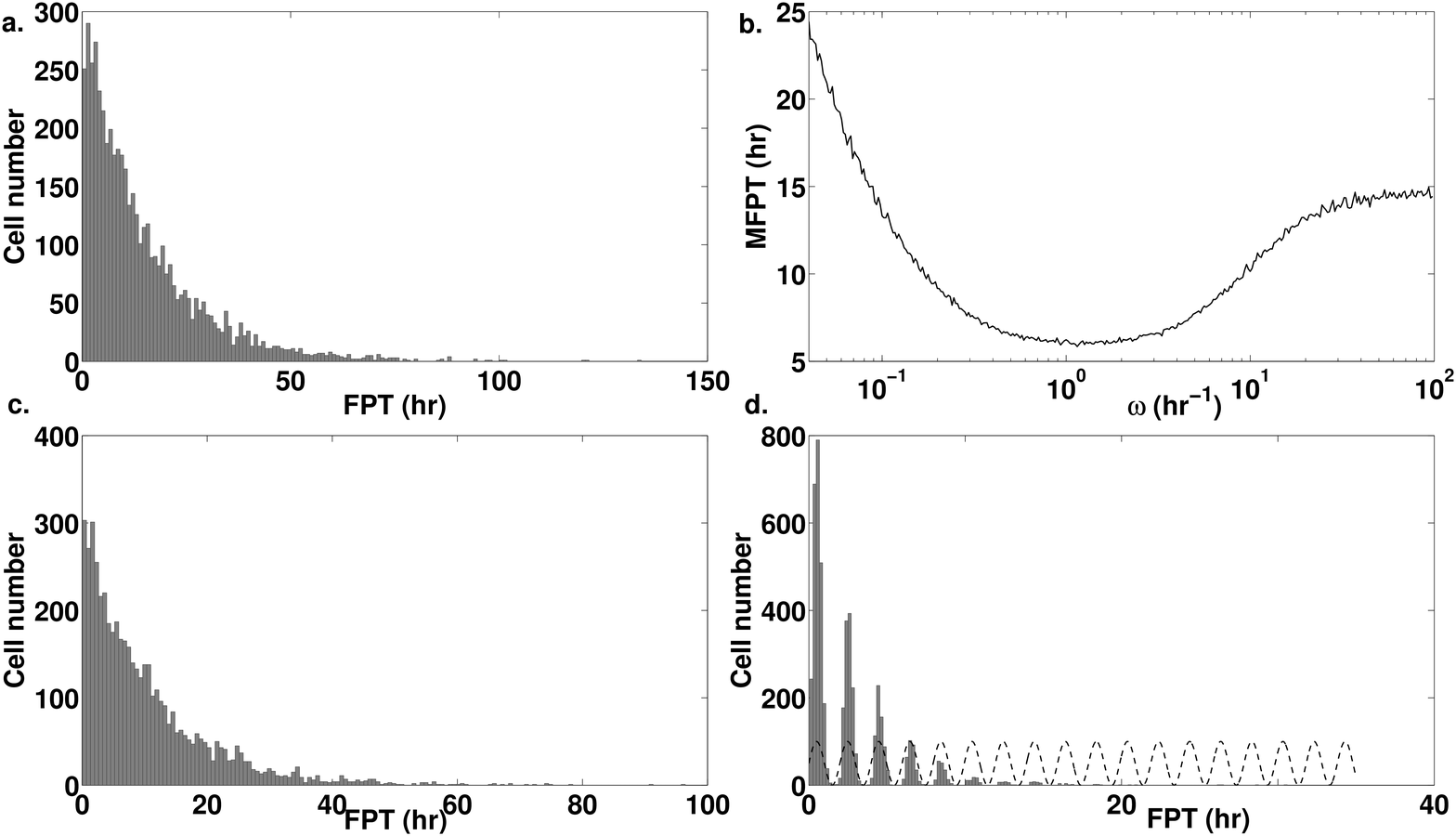}}
 \caption{
 \label{Figure2}}
\end{figure}

\begin{figure}
 \centerline{\includegraphics[width=100mm]{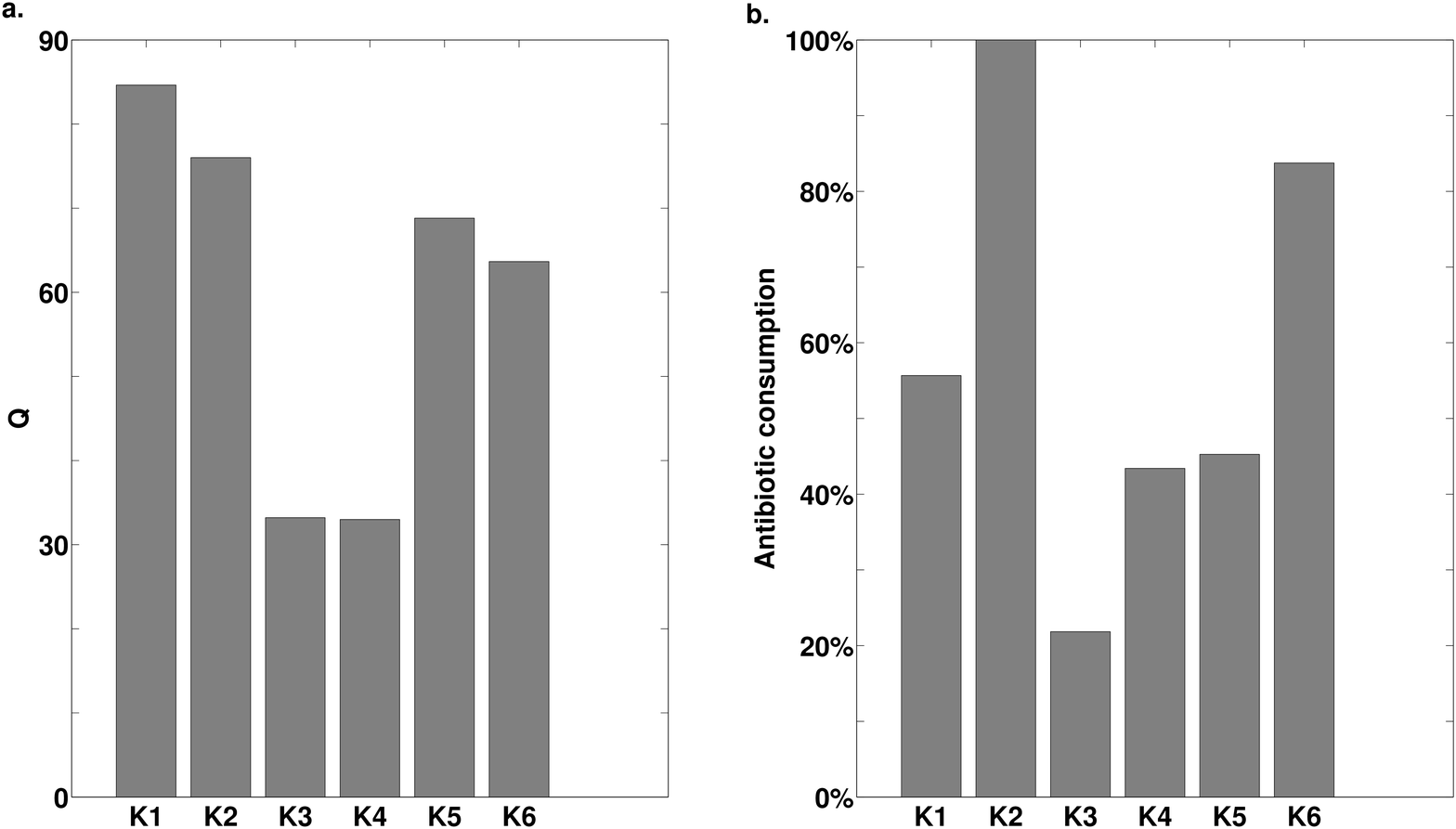}}
 \caption{
 \label{Figure3}}
\end{figure}

\begin{figure}
 \centerline{\includegraphics[width=100mm]{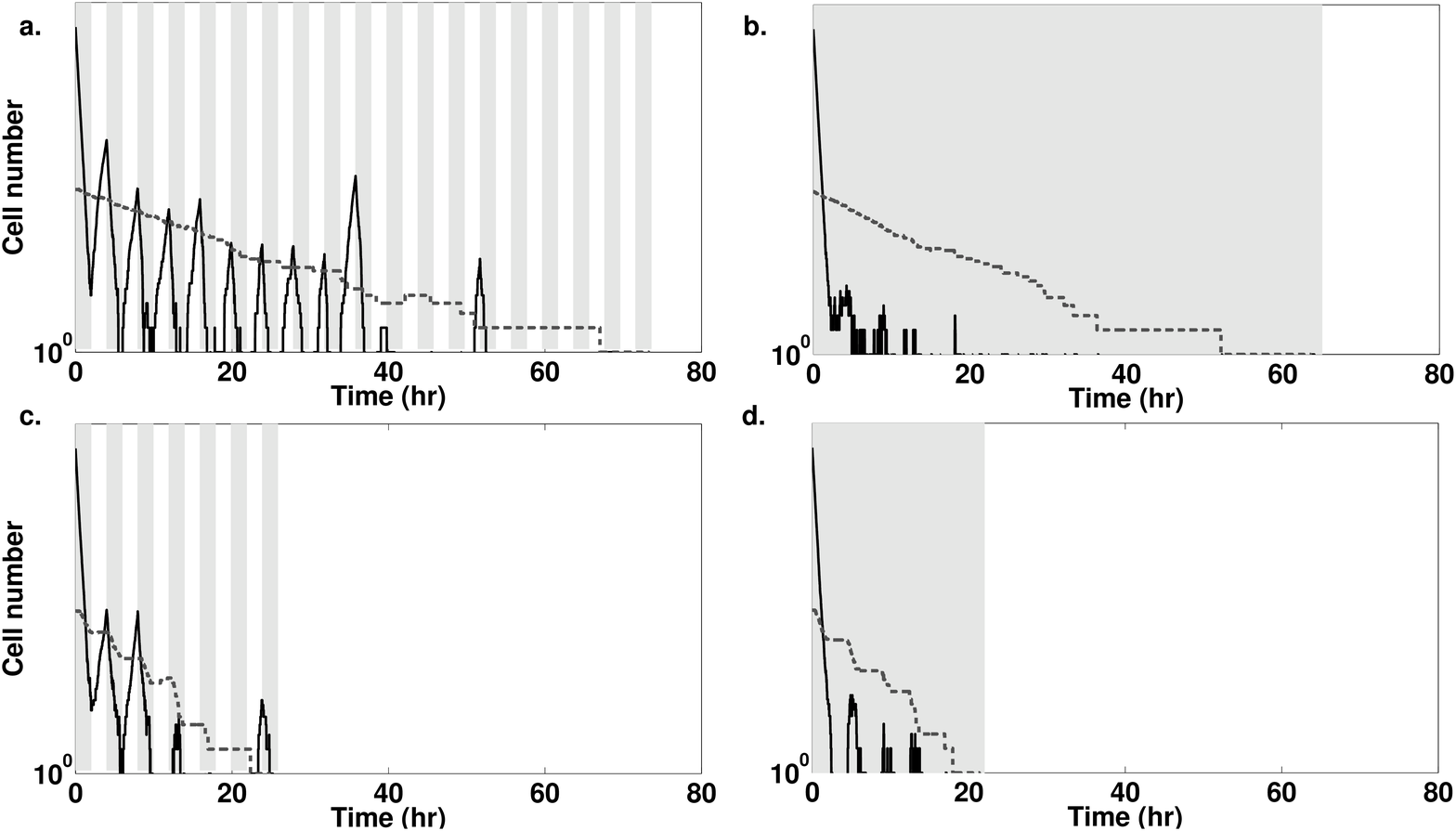}}
 \caption{
 \label{Figure4}}
\end{figure}

\begin{figure}
 \centerline{\includegraphics[width=100mm]{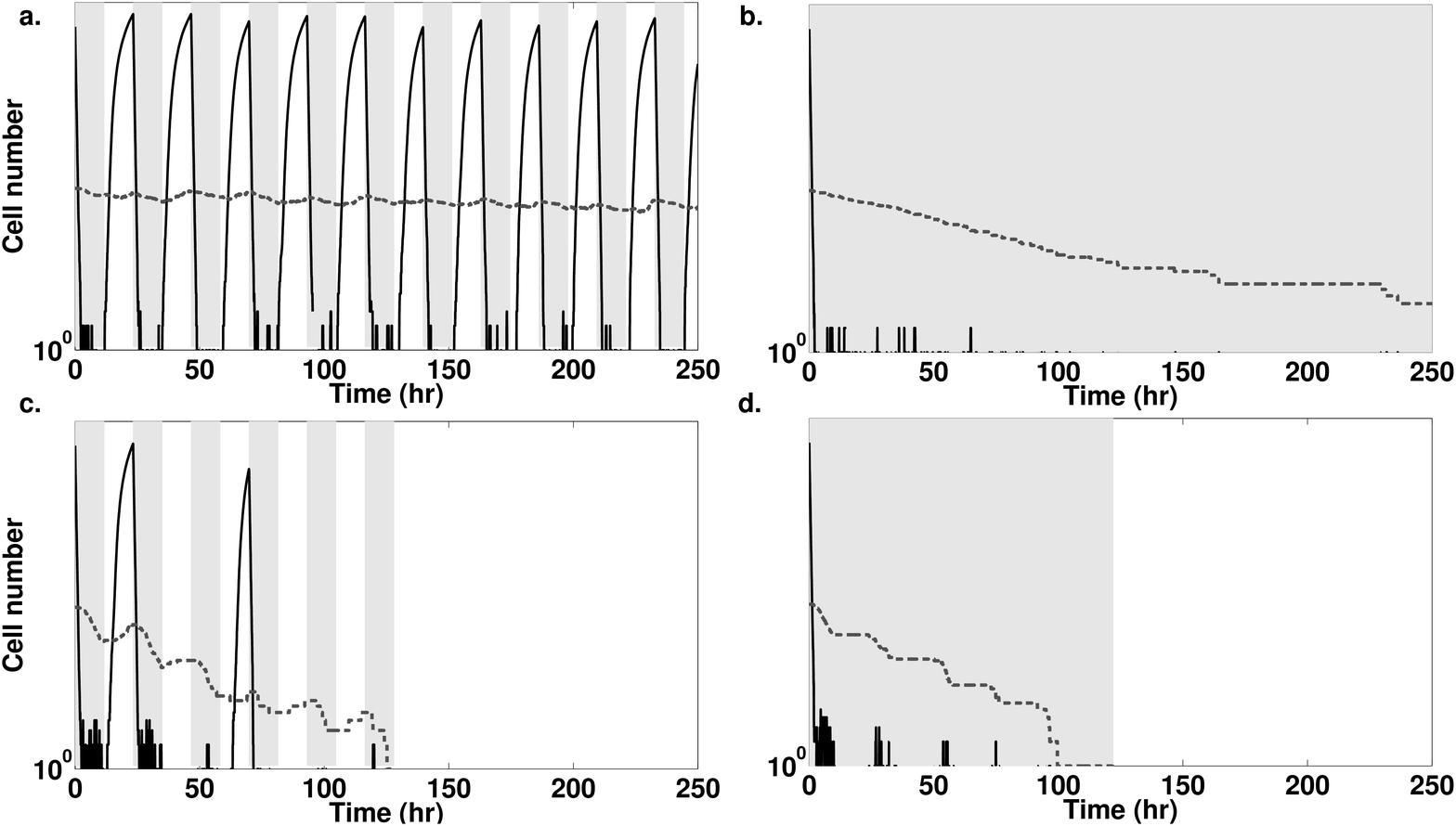}}
 \caption{
 \label{Figure5}}
\end{figure}

\clearpage
\section*{Figure Legends.}
\itemize
\item Figure.1 A schematic toggle switch model controlling single cell phenotype switch. (a) The toggle switch model. Periodic signal is added to the system, perturbing the degradation rate of protein A. (b)Gilliespie simulations show that the model behaves like a two-state system, corresponding to two phenotypes of bacterium.\\
\item Figure 2. Results of single cell simulations. (a) $P(t_{p2n})$ in the absence of the perturbing signal. (b) The mean first passage time (MFPT) versus frequency $\omega$ of the sine-formed perturbation signal. (c) A weak perturbing signal with off-resonance frequency still gives an exponential distribution of $P(t_{p2n})$, here $\omega=10 hr^{-1}$. (d) A weak perturbing signal with resonance frequency (red curve) changes $P(t_{p2n})$ into several separated spikes. Each spike overlaps with a peak of the periodic signal (red curve). $\omega_c=1.6 hr^{-1}$. \\
\item Figure 3. Comparison of various strategies at the population level. (a)  90\% quantile ($Q$) of $T_{kill}$ under six different killing strategies. (b) Corresponding relative antibiotic consumption.  Please refer to Table 1 for the illustration of killing strategies K.\\
\item Figure 4. Sample population dynamics corresponding to the results in Figure 3. (a) K1. (b) K2. (c) K3, $\omega_c=1.6\  hr^{-1}$, $T_{\omega_c}=2\pi/{\omega_c}=3.9\ hr$. (d) K4.  Black curve shows the dynamics of normally growing population. Red curve shows the dynamics of persister population. Gray time windows in the background indicate antibiotic treatment, blank time windows represent environment good for growth.\\
\item Figure 5. Sample population dynamics corresponding to a phenotype with slower switching rate (see the main text). Antibiotic treatment is either continuous or periodic. (a) K1.  (b) K2.  (c) K3, $\omega_c=0.27\  hr^{-1}$, $T_{\omega_c}=2\pi/{\omega_c}=23.3\ hr$. (d) K4.  Black curve shows the dynamics of normally growing population. Red curve shows the dynamics of persister population. Gray time windows in the background indicate antibiotic treatment, blank time windows represent environment good for growth. \\

\clearpage
\begin{table}[ht]
 \caption{Parameters for single cell level simulation}
 \centering
 \begin{tabular}{c c c}
 \hline\hline
    parameter & value (reduced unit $^*$) & notation\\
    \hline
   $k_1^0$ & 0.05 & \parbox[t]{6cm}{\raggedright basal synthesis rate of A}\\
   $k_2^0$ & 1 & \parbox[t]{6cm}{\raggedright basal synthesis rate of B }\\
   $k_3^0$ & 0.45 & \parbox[t]{6cm}{\raggedright basal degradation rate of A} \\
   $k_4^0$ & 0.56 & \parbox[t]{6cm}{\raggedright basal degradation rate of B} \\
   $k_5$ & 5 & \parbox[t]{6cm}{\raggedright dimmer $A_2$ association rate}\\
   $k_{-5}$ & 5 & \parbox[t]{6cm}{\raggedright dimmer $A_2$ disassociation rate}\\
   $k_6$ & 5 & \parbox[t]{6cm}{\raggedright dimmer $B_2$ association rate}\\
   $k_{-6}$ & 5 & \parbox[t]{6cm}{\raggedright dimmer $B_2$ dissociation rate}\\
   $k_7$ & 5 & \parbox[t]{6cm}{\raggedright binding rate between operon and $A_2$}\\
   $k_{-7}$ & 1 & \parbox[t]{6cm}{\raggedright unbinding rate of $A_2$ from operon}\\
   $k_8$ & 5 & \parbox[t]{6cm}{\raggedright binding rate between operon and $B_2$}\\
   $k_{-8}$ & 1 & \parbox[t]{6cm}{\raggedright unbinding rate of $B_2$ from operon}\\
   $k_9$ & 1 & \parbox[t]{6cm}{\raggedright synthesis rate of A}\\
   $k_{10}$ & 1 & \parbox[t]{6cm}{\raggedright synthesis rate of B}\\
    \hline
 \end{tabular}
 *: $unit = 130\ hr^{-1}$
\end{table}

\clearpage
\begin{table}[ht]
 \caption{Parameters for population level simulation}
 \centering
 \begin{tabular}{c  c  c}
 \hline\hline
    parameter & value ($hr^{-1}$) & notation\\
    \hline
    $g_n$ & 0.2  & \parbox[t]{7cm}{\raggedright net growth rate of normally growing cells under growth condition}\\
    $d_n$ & 4  & \parbox[t]{7cm}{\raggedright net death rate of normally growing cells under antibiotic condition}\\
    $g_p$ & 0.02  & \parbox[t]{7cm}{\raggedright net growth rate of persister cells under growth condition}\\
    $d_p$ & $1\times 10^{-6}$  & \parbox[t]{7cm}{\raggedright net death rate of persister cells under antibiotic condition}\\
       \hline
 \end{tabular}
\end{table}

\clearpage
\begin{table}[ht]
 \caption{90\% quantile of $T_{kill}$ ($Q$) under different killing strategies}
 \centering
 \begin{tabular}{c |c  c c  c}
 \hline\hline
    \ & $S_{\omega_1}$ & $E_{\omega_2}$ & $Q\ (hrs)$ & notation\\
    \hline
    $K_1$ & $\phi$ & $H(S_{\omega_c})^{*}$ & 84.6 &\parbox[t]{9cm}{\raggedright no signal + periodic antibiotics} \\ 
    $K_2$ & $\phi$ & $E_0^{**}$ & 76.0 &\parbox[t]{9cm}{\raggedright no signal + continuous antibiotics} \\ 
    $K_3$ & $S_{\omega_c}$ & $H(S_{\omega_c})$ & 33.2 &\parbox[t]{9cm}{\raggedright resonance signal + periodic antibiotics}  \\ 
    $K_4$ & $S_{\omega_c}$ & $E_0$ &33.0 & \parbox[t]{9cm}{\raggedright resonance signal +  continuous antibiotics}\\ 
    $K_5$ & $S_{10}$ & $H(S_{10})$ & 68.8&\parbox[t]{9cm}{\raggedright off-resonance signal + off-resonance periodic antibiotics}\\ 
    $K_6$ & $S_{10}$ & $E_0$ &63.7& \parbox[t]{9cm}{\raggedright off-resonance signal + continuous antibiotics}\\ 
    \hline
 \end{tabular}
\raggedright *: $H(x)$ is a Heaviside function which returns 1 if $x\ge 0$, otherwise returns 0. We assume $H=1$ denotes the antibiotic treatment is switched on, and $H=0$ denotes the antibiotic treatment is switched off.  $S_{\omega_c}=Asin(\omega_ct)$ is the sine-formed perturbing signal under resonance frequency $\omega_c$. **: $E_0$ represents continuous antibiotic treatment (the antibiotic treatment is always kept on).
\end{table}

\clearpage
\section{Supplementary Information}

\subsection{Large region of $\omega$ guarantees RA in single cell phenotypic transition }
\begin{figure}
 \centerline{\includegraphics[width=140mm]{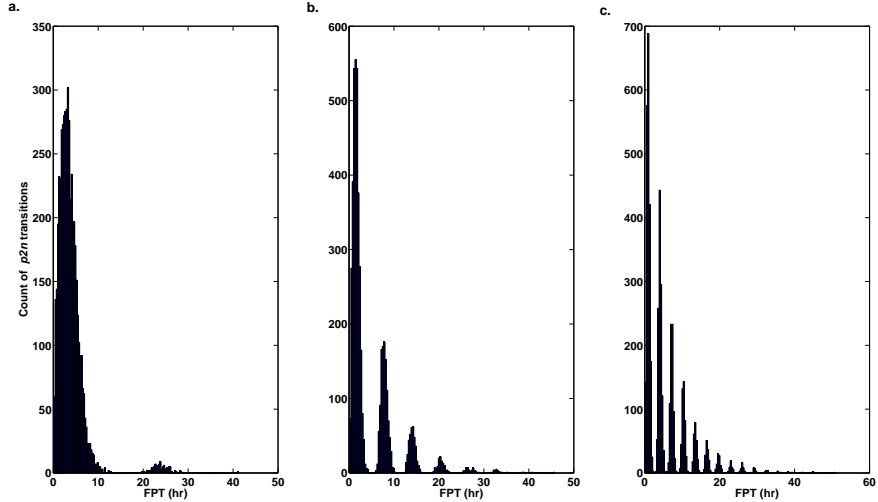}}
 \caption{$P(t_{p2n})$ under different signal frequency. All signals are sine-formed, as described in the main text. a. $\omega = 0.3\ hr^{-1}$.  b. $\omega = 1.0\ hr^{-1}$. c. $\omega = 2.0\ hr^{-1}$. All these three cases give values of the MFPT close to the minimum in Figure 2 (b) in the main text.
 \label{Figure6}}
\end{figure}
Under resonance frequency $\omega_c\approx 1.6\ hr^{-1}$, the mean first passage time (MFPT) reaches a minimum. Actually, the MFPT-$\omega$ curve displays a rather flat bottom, indicating a large region of of the signal frequency may generate RA in our system. Figure 6 shows some FPT distribution for the {\it p2n} transitions (also denoted by $P(t_{p2n})$) within this region of signal frequency.  The $P(t_{p2n})$ under $\omega=0.3, 1.0, 2.0 \ hr^{-1}$ all produce separated-spikes-like distribution. This allows some freedom in real therapeutical applications. We have repeated the double well system studied by Schmitt {\it et al.} \cite{Schmitt2006}, and found similar behaviors.

\subsection{Choices of the perturbation target for RA}
\begin{figure}
 \centerline{\includegraphics[width=140mm]{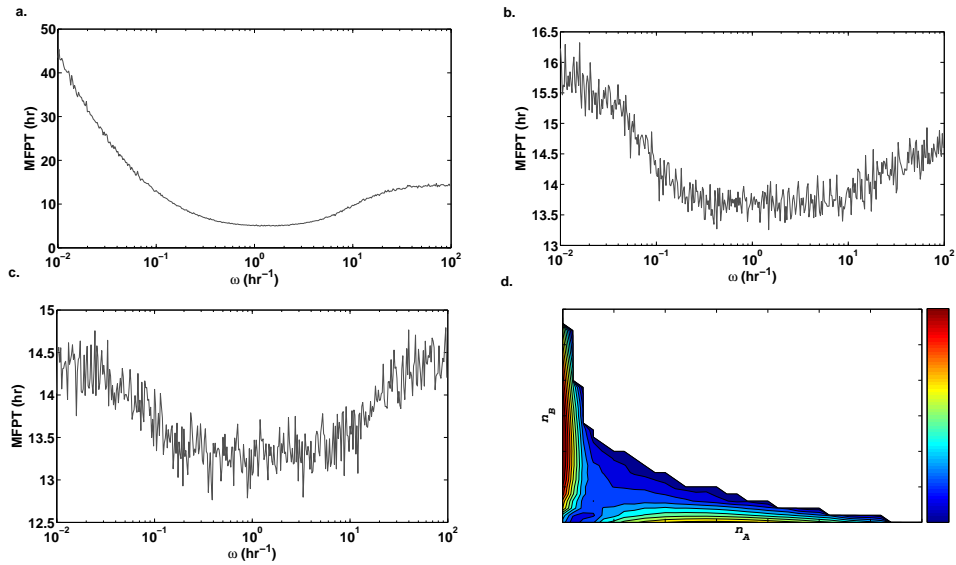}}
 \caption{RA with perturbation on different reactions. a) MFPT of $p2n$ transition when A's synthesis rate $k_9$ is perturbed. b) MFPT of $p2n$ transition when B's synthesis rate $k_{10}$ is perturbed. c) MFPT of $p2n$ transition when B's degradation rate $k_4^{0}$ is perturbed. These perturbations are implemented using sine signal with the same perturbing strength.  Together with the case of perturbing A's degradation, all four cases generate local minimums around $w_c=1.6 hr^{-1}$, but with different fold of change in the MFPT as signal frequency varies. d) Stationary distribution on the $n_A-n_B$ phase plane sampled with a long trajectory. Refer to the colorbar for the relative probability of each ($n_A$,$n_B$), with a decreasing value from the top (red) to the bottom (blue). 
 \label{Figure7}}
\end{figure}

In the main text, we reported results with the degradation rate of protein A being periodically perturbed. Figure 7 (a-c)  shows  the MFPT of $p2n$ transitions when A's synthesis rate, B's synthesis or degradation rates are perturbed at different signal frequency, respectively. Clearly,  resonant activation exists in all cases.  However, the fold of change between the MFPT at the resonant frequency and that with no perturbation varies. The effect with perturbation on A's synthesis is similar to that on A's degradation, but the effect with either B's synthesis or degradation is much weaker. Figure 7 (d) shows the system's stationary distribution on the $n_A-n_B$ plane, which can be related to a potential \cite{Graham1971}. Clearly the system dynamic shows transition state-like behavior, with the transition state having small numbers of both A and B. That is, leaving from the state with high A and low B (persister), for the system to make a state transition it is more important to reduce the number of A than increasing the number of B. This explains why it is more sensitive to perturb the reactions involving A.

\subsection{Resonant activation exists with various signal forms}
\begin{figure}
 \centerline{\includegraphics[width=140mm]{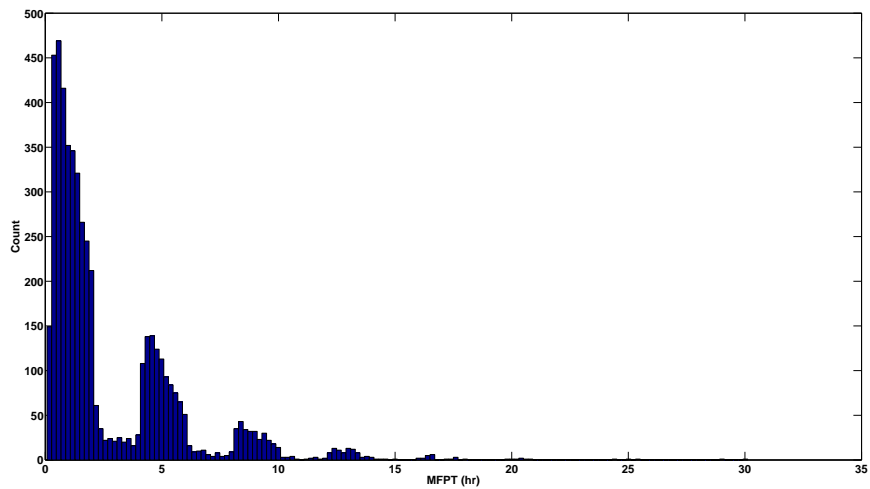}}
 \caption{$t_{p2n}$ under the step-function signal. Other parameters are the same as in Table 1 of the main text.
 \label{Figure8}}
\end{figure}
In the main text, we applied a sine-formed signal to perturb the degradation rate of protein A in each generic toggle switch, and observed RA. A more practical signal form may be unidirectional on the perturbation. We tested with a step-function form. That is, the protein degradation rate $k_3$ was only increased from its basal rate periodically by a constant value. Again we observed that the MPFT shows a minimum at the same $\omega_c$ as with the sine function form. However, in this case some of  the neighboring spikes in the FPT distribution are  not fully separated, Figure 8. We observed $\sim 9.4\%$ $p2n$ transitions taking place under the growth environment, comparing to $\sim 4.3\%$ for the sine function form. 
Consequently, for a system with easy population restoration ({\it i.e}., the second example discussed in the main text), the sterilization time using the periodic antibiotics treatment  (K3) with the step function signal  is longer than that with the sine function signal. In this case, strategy K4 under the step function signal gives a $90\%$ quantile of the  sterilization time $\sim 39\%$ of that with strategy K2.

\subsection{The distribution of $T_{kill}$ and the reason for using statistics $Q$}

\begin{figure}
 \centerline{\includegraphics[width=140mm]{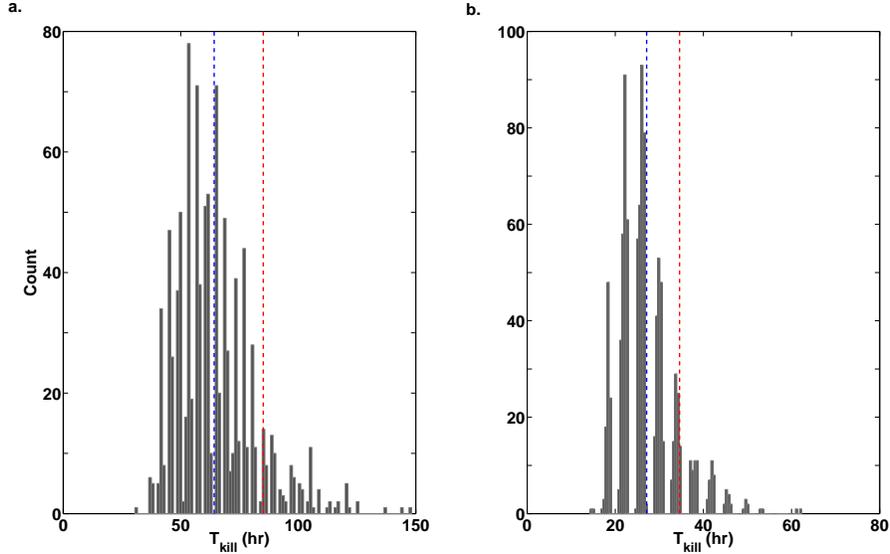}}
 \caption{Distribution of $T_{kill}$.  a. Case K1. b. Case K3. Blue dash line shows the location of the mean value; red dash line shows the location of 90\% quantile of the total samples ($Q$).
 \label{Figure9}}
\end{figure}

Figure 9 gives the killing time distribution of the 1000 samples under  strategies K1 and K3. Clearly both distributions have long tails. Therefore we suggest that the 90\% quantile may be a better quantity than the mean value to compare different strategies, although the conclusions are the same.

\subsection{Most $p2n$ transitions take place within antibiotic periods under killing strategy K3}
\begin{figure}
 \centerline{\includegraphics[width=140mm]{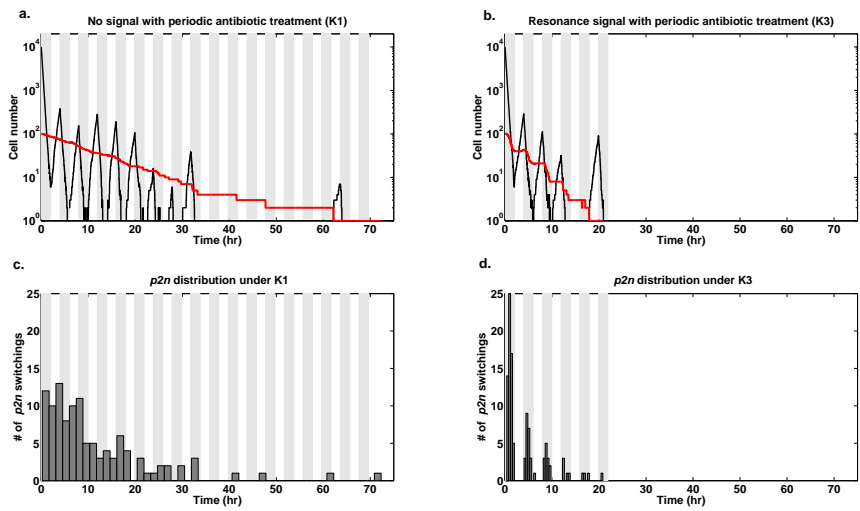}}
 \caption{Sample population dynamics with corresponding $p2n$ transitions under different killing strategies. (a) K1. (b) K3, $\omega_c=1.6\  hr^{-1}$, $T_{\omega_c}=2\pi/{\omega_c}=3.9\ hr$. (c) $p2n$ under K1.  (d) $p2n$ under K3.  Black curve shows the dynamics of normally growing population. Red curve shows the dynamics of persister population. Gray time windows in the background indicate antibiotic treatment, blank time windows represent environment good for growth.
 \label{Figure10}}
\end{figure}
By recording the population dynamics with corresponding $p2n$ transitions under different killing strategies, Figure 10 shows that most $p2n$ transitions take place within the period of antibiotics treatment under strategy K3. Therefore, the bacterial population is difficult to re-establish when K3 is applied. On the other hand, under K1, the distribution of $p2n$ is exponential, indicating the phenotypic transition may take place during either growth condition or antibiotic condition. 

\subsection{Signal strength affects the sterilization time of bacteria population}
\begin{figure}
\centerline{\includegraphics[width=140mm]{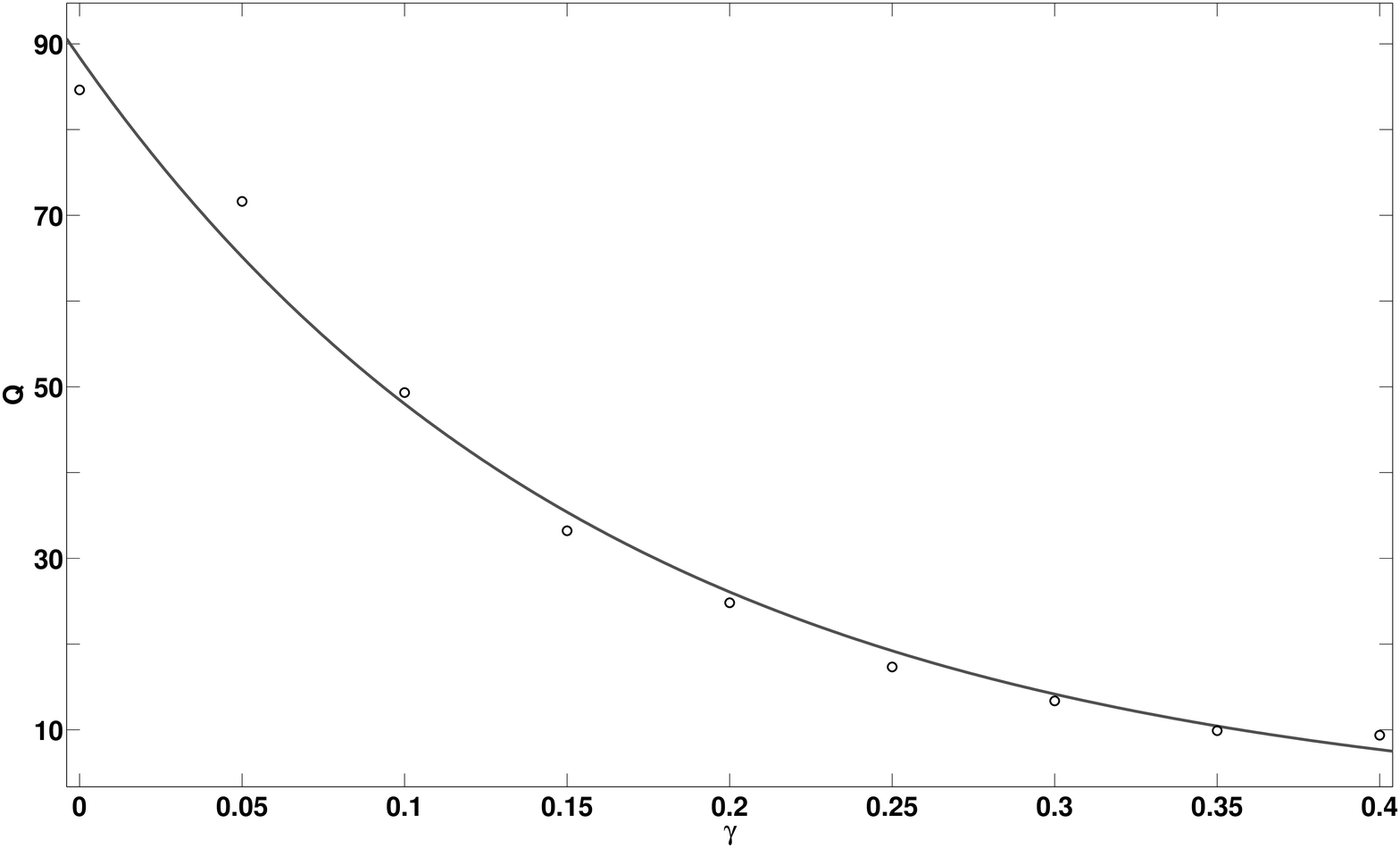}}
 \caption{90\% quantile of sterilization time ($Q$) under signal with different strength. Here we use sine-formed signal $S=\gamma sin(\omega t)$ with strength $\gamma$. In the main text, $\gamma=0.15 \ unit$. The maximal $\gamma$ should be the basal degradation rate of the target protein, which in our case is $0.45 \ unit$.
 \label{Figure11}}
\end{figure}
As can be seen from Figure S11, signals with stronger strength can reduce more time needed for sterilizing bacteria population. However, a stronger signal implies a bigger perturbation to the system, which may be unfavorable due to practical considerations and toxicity.

\subsection{Duration of each session of antibiotics treatment affects sterilization time}
\begin{figure}
 \centerline{\includegraphics[width=140mm]{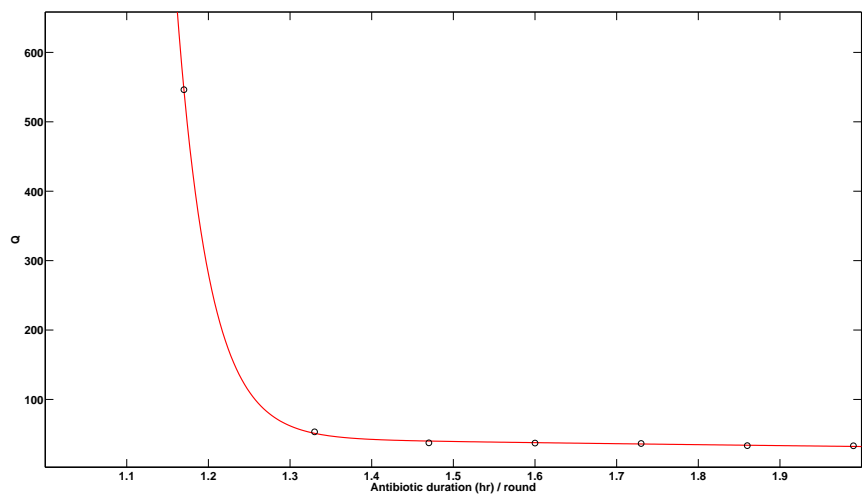}}
 \caption{90\% quantile of sterilization time ($Q$) under different duration of antibiotic treatment. 
 \label{Figure12}} 
\end{figure}

At resonance frequency,  the $p2n$ transition distribution is multi-spike shaped. This allows discrete antibiotics treatment, as long as each session covers all or most of the $p2n$ transition time.  We performed population level simulations with different antibiotics treatment duration centered at the signal peak times. Figure 12 shows the 90\% quantile of the sterilization time. As expected, upon increasing the antibiotics treatment duration from zero, the sterilization time first drops sharply,  and then flattens.  Further increasement of the antibiotic duration does not decrease the sterilization time. Therefore in this case one may further reduce the duration of each session of the antibiotics treatment to $\sim 1.3\  hr$ from the $2 hr$ value used in in the main text, without significant increase of the sterilization time.

\end{document}